\documentclass[12pt,preprint]{aastex}

\usepackage{CJK}
\usepackage{natbib}
\usepackage{graphicx} 
\usepackage{epsfig}

\begin{document}
\begin{CJK}{UTF8}{gbsn}

\title{The Drop during Less than 300 days of a Dusty White Dwarf's Infrared Luminosity}

\author{S. Xu(许\CJKfamily{bsmi}偲\CJKfamily{gbsn}艺)\altaffilmark{a}, M. Jura\altaffilmark{a}}
\altaffiltext{a}{Department of Physics and Astronomy, University of California, Los Angeles CA 90095-1562; sxu@astro.ucla.edu, jura@astro.ucla.edu}

\begin{abstract}
We report Spitzer/IRAC photometry of WD J0959$-$0200, a white dwarf that displays excess infrared radiation from a disk, likely produced by a tidally disrupted planetesimal. We find that in 2010, the fluxes in both 3.6 $\mu$m and 4.5 $\mu$m decreased $\sim$ 35\% in less than 300 days. The drop in the infrared luminosity is likely due to an increase of the inner disk radius from one of two scenarios: (i) a recent planetesimal impact; (ii) instability in the circumstellar disk. The current situation is tantalizing; high sensitivity, high cadence infrared studies will be a new tool to study the interplay between a disk and its host white dwarf star.

\end{abstract}


\section{Introduction}

The discovery of an infrared excess around a single white dwarf G29-38 \citep{ZuckermanBecklin1987} from an orbiting dust disk \citep{Graham1990b} has helped lay the foundation for the study of planetary systems around white dwarfs. However, it was not until much later that a now widely-accepted model was proposed to explain the source of the dust -- tidal disruption of planetesimals \citep{DebesSigurdsson2002,Jura2003}. Subsequent progress has been rapid, particularly after the discovery of an infrared excess around a second white dwarf GD 362 \citep{Kilic2005, Becklin2005}. Various dynamical pathways have been explored to deliver the planetesimals into the white dwarf tidal radius  \citep{Bonsor2011, Veras2013, FrewenHansen2014, Mustill2014, Veras2013}.  Eventually, all these materials get accreted onto the white dwarf and enrich its pure hydrogen or/and helium atmosphere. Studying these heavy-element-enriched white dwarfs has become a uniquely powerful way to directly measure the bulk compositions of extrasolar planetesimals \citep{vonHippel2007, Zuckerman2007, Klein2010, Farihi2010b, Gaensicke2012, Jura2012, Xu2014, JuraYoung2014}.

To-date, mainly thanks to the launch of the Spitzer Space Telescope, more than 30 dusty white dwarfs have been identified \citep{Farihi2009, Debes2011b, XuJura2012,Barber2014}. The SEDs can be fit with a geomterically thin, optically thick disk within the tidal radius of the white dwarf \citep{Jura2003}. The typical disk lifetime is estimated to be 10$^5$ - 10$^6$ yr \citep{Jura2008, Rafikov2011a, Rafikov2011b, Girven2012}. The disk occurrence rate is about 4\% for white dwarfs between 9500-22,500 K \citep{Barber2012}. Despite intensive searches, only one disk host star G166-58 has an effective temperature less than 9500 K \citep{Farihi2008b, XuJura2012}. When observed with a high-resolution optical/ultraviolet spectrograph, all dusty white dwarfs show a high level of atmospheric enrichment of heavy elements\footnote{With the Wide-field Infrared Survey Explorer (WISE), \citet{Hoard2013} identified three potential dusty white dwarfs with no heavy elements in their atmospheres. However, they also noted that they are only candidates for having a dust disk due to possible background contamination.}. Circumstellar gaseous material has also been detected around some white dwarfs \citep{Gaensicke2006, Melis2010, Brinkworth2012, Debes2012b}.

There are a lot of puzzles regarding these dusty white dwarfs.  How massive is the dust disk? How does the disk evolve with time? Why do some of them show calcium triplet emission lines from the circumstellar gas while others do not? In addition, these emission lines appear to be variable on timescale of a few years \citep{Gaensicke2008}. What causes such variability?

In this paper, we report the first result from a systematic study of infrared variability for dusty white dwarfs. We focus on one target, WD J0959$-$0200, whose fluxes have decreased by $\sim$ 35\% in the infrared.

\section{Observations and Data Reduction}

\subsection{Source Selection}

Most dusty white dwarfs have been observed with the Infrared Array Camera (IRAC) \citep{Fazio2004} at 3-10 $\mu$m, where the disk emits most of its flux. When the WISE data became available, we compared the fluxes of all dusty white dwarfs at W-1 with IRAC-1 and W-2 with IRAC-2. We found that the flux level was comparable for the majority of them but a few showed disparities larger than 3 sigma; most of them have a larger WISE fluxes due to the lower spatial resolution of WISE and contamination from background sources (also see Figure \ref{Fig: Image}). In particular, five white dwarfs were in a clean field with no obvious contaminating source in the IRAC image and this discrepancy can be real. We were awarded Spitzer time in cycle 10 to re-investigate these five targets. The program is still on-going and we report the result of WD J0959$-$0200 in this paper. 

WD J0959$-$0200 is a hydrogen-dominated white dwarf discovered from the Sloan Digital Sky Survey (SDSS) with a K band excess from the UKIRT (United Kingdom Infrared Telescope) Infrared Deep Sky Survey (UKIDSS) \citep{Girven2011}. Follow-up observation from Spitzer shows that this star indeed has an infrared excess from an orbiting dust disk. Spectroscopic observations with the Very Large Telescope (VLT) show photospheric absorption lines from calcium, magnesium as well as infrared calcium triplet emission \citep{Farihi2012a}. With a stellar temperature of 13,280 K, WD J0959$-$0200 is the coolest dusty white dwarf with circumstellar gaseous material, whose effective temperatures range between 13,280 K and 20,700 K. In comparison, dusty white dwarfs typically have a temperature range of 9500 - 23,100 K \citep{Brinkworth2012}.

Because a large portion of the K band flux for WD J0959$-$0200 is also from the dust disk, we obtained additional JHK photometry with the UKIRT. The observing logs are presented in Table \ref{Tab: Photometry}. 

\subsection{Spitzer}

WD J0959$-$0200 was observed with IRAC on warm Spitzer \citep{Werner2004} in 2010 (PI: J. Farihi). The observing strategy and data analysis were described in \citet{Farihi2012a}. We reobserved this target in 2014 (PI: M. Jura) with a similar configuration -- 30 sec exposure with 30 medium size dithers in the cycling pattern, resulting in a total of 900 sec on target time in each channel.

\begin{figure}[hp]
\centering
\begin{minipage}{.3\textwidth}
\centering
\includegraphics[width=\linewidth]{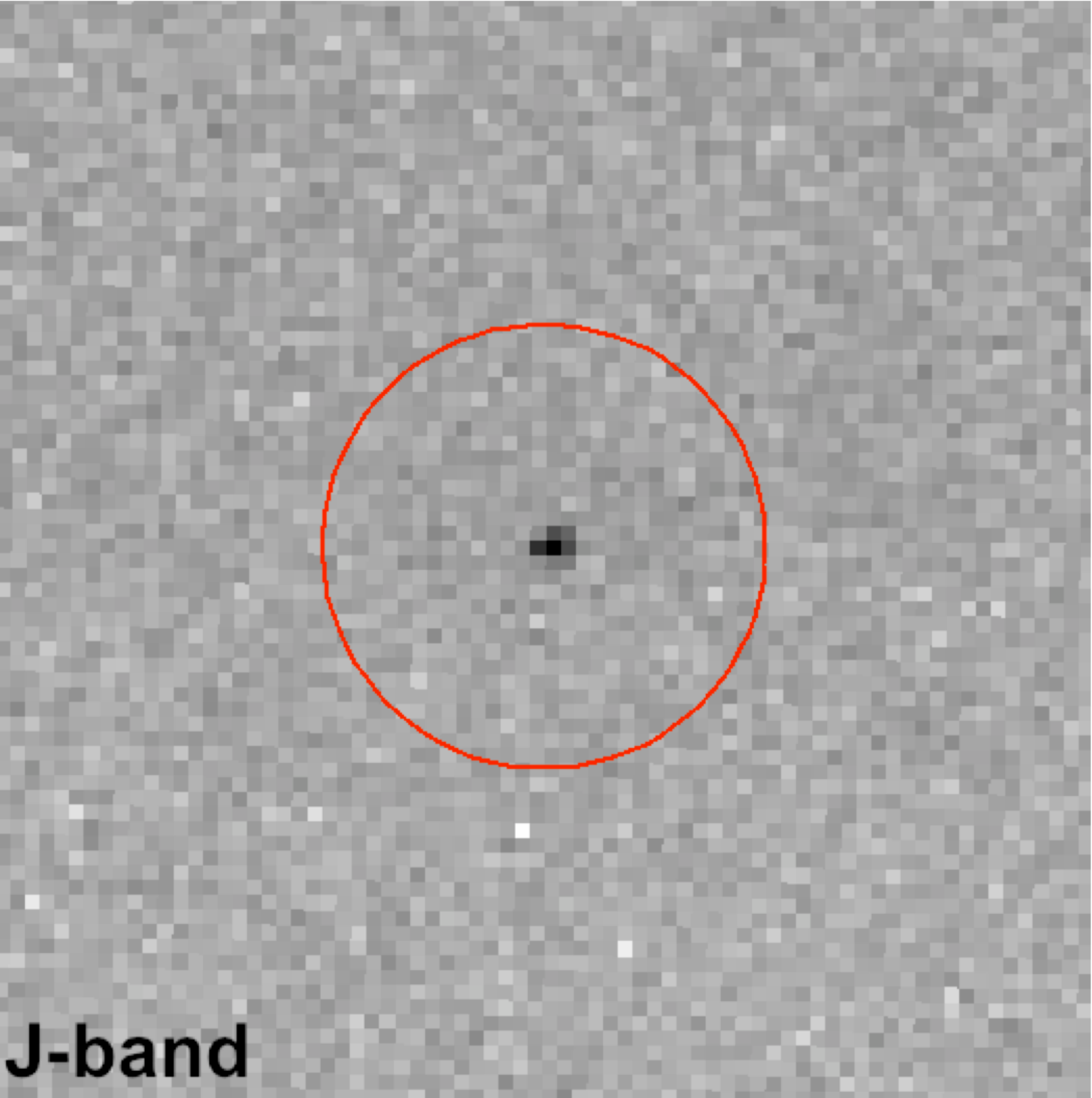}
\end{minipage}\hfill
\begin{minipage}{.3\textwidth}
\centering
\includegraphics[width=\linewidth]{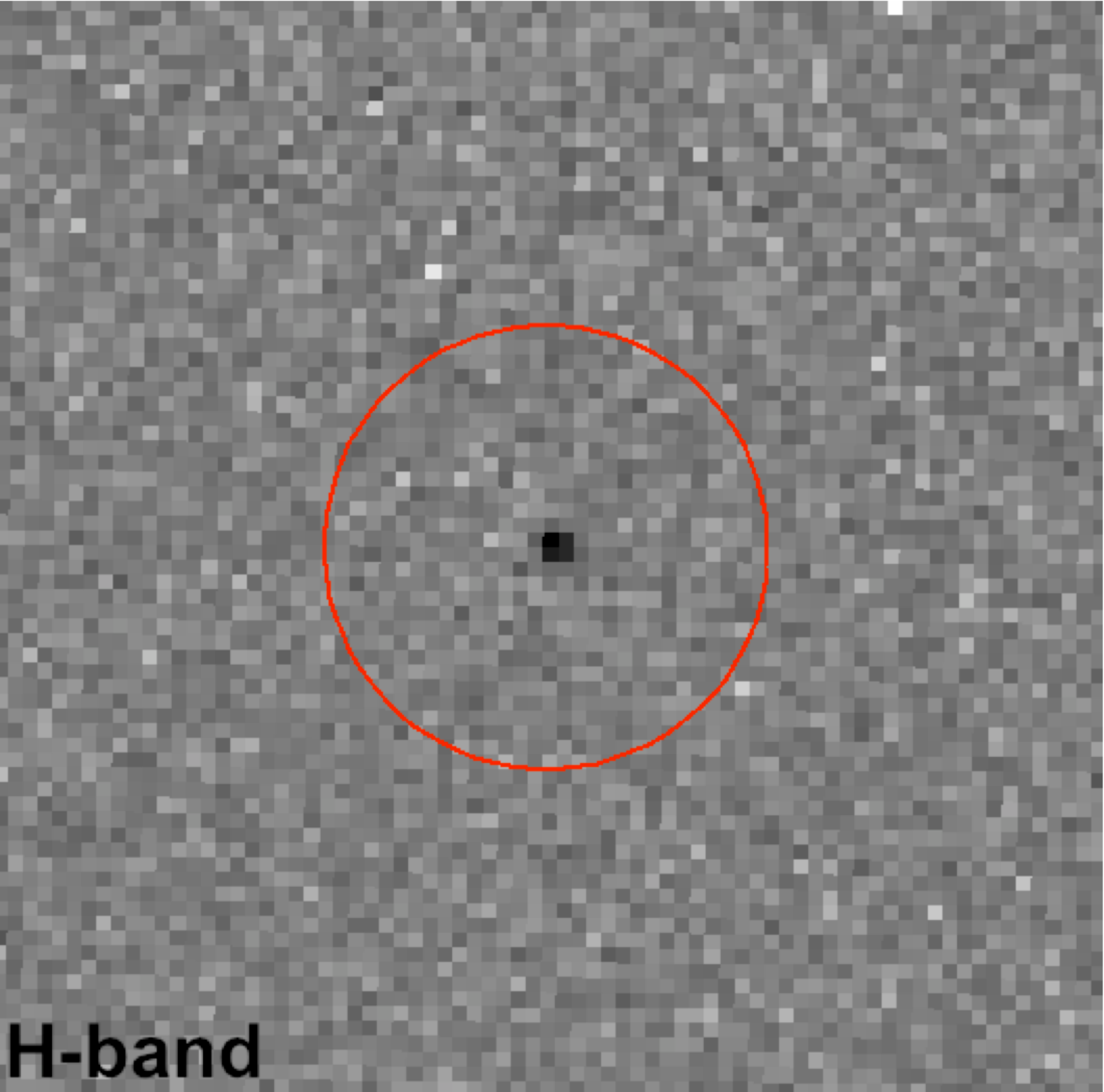}
\end{minipage}\hfill
\begin{minipage}{.3\textwidth}
\centering
\includegraphics[width=\linewidth]{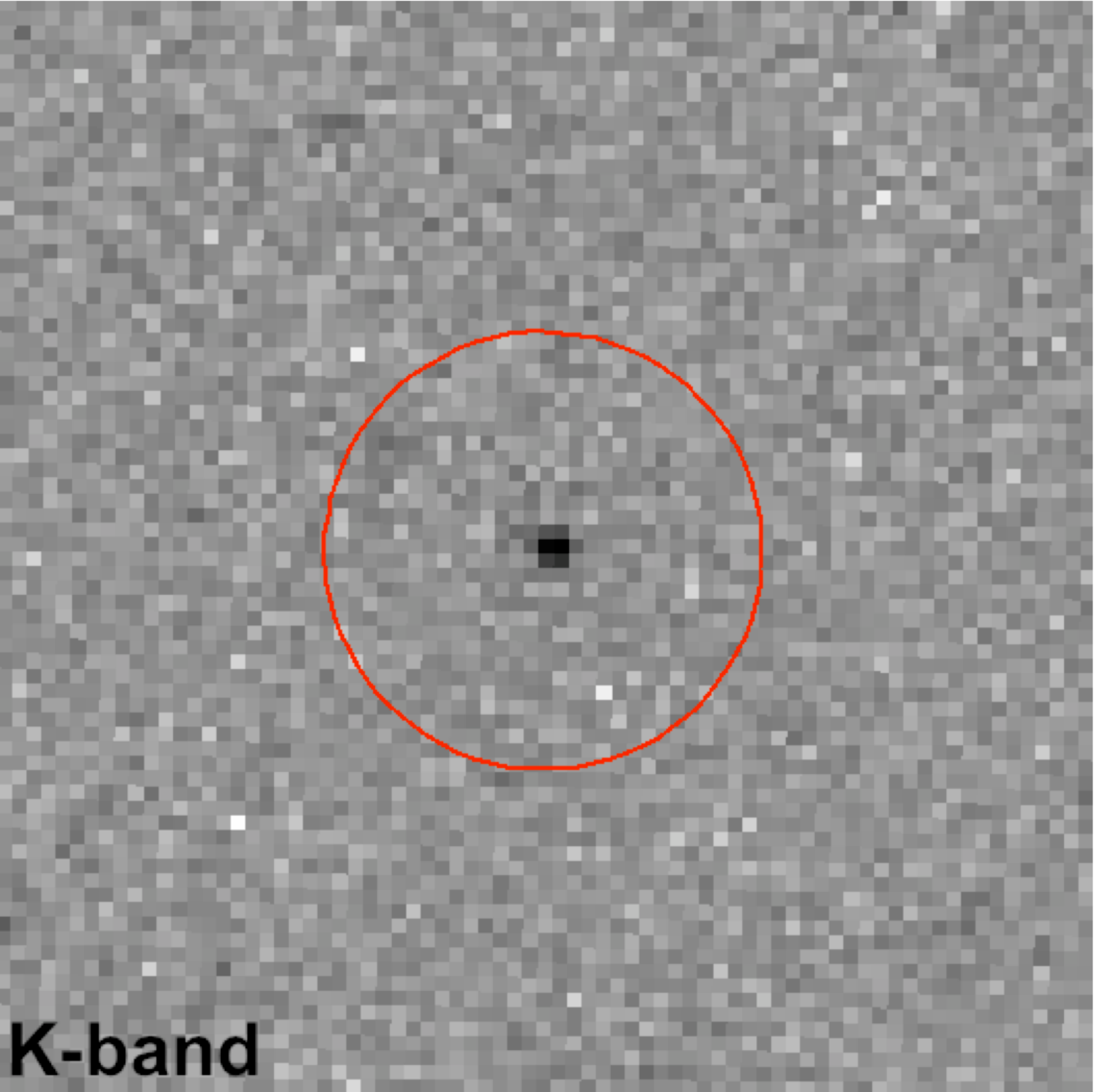}
\end{minipage}
\begin{minipage}{.3\textwidth}
\centering
\includegraphics[width=\linewidth]{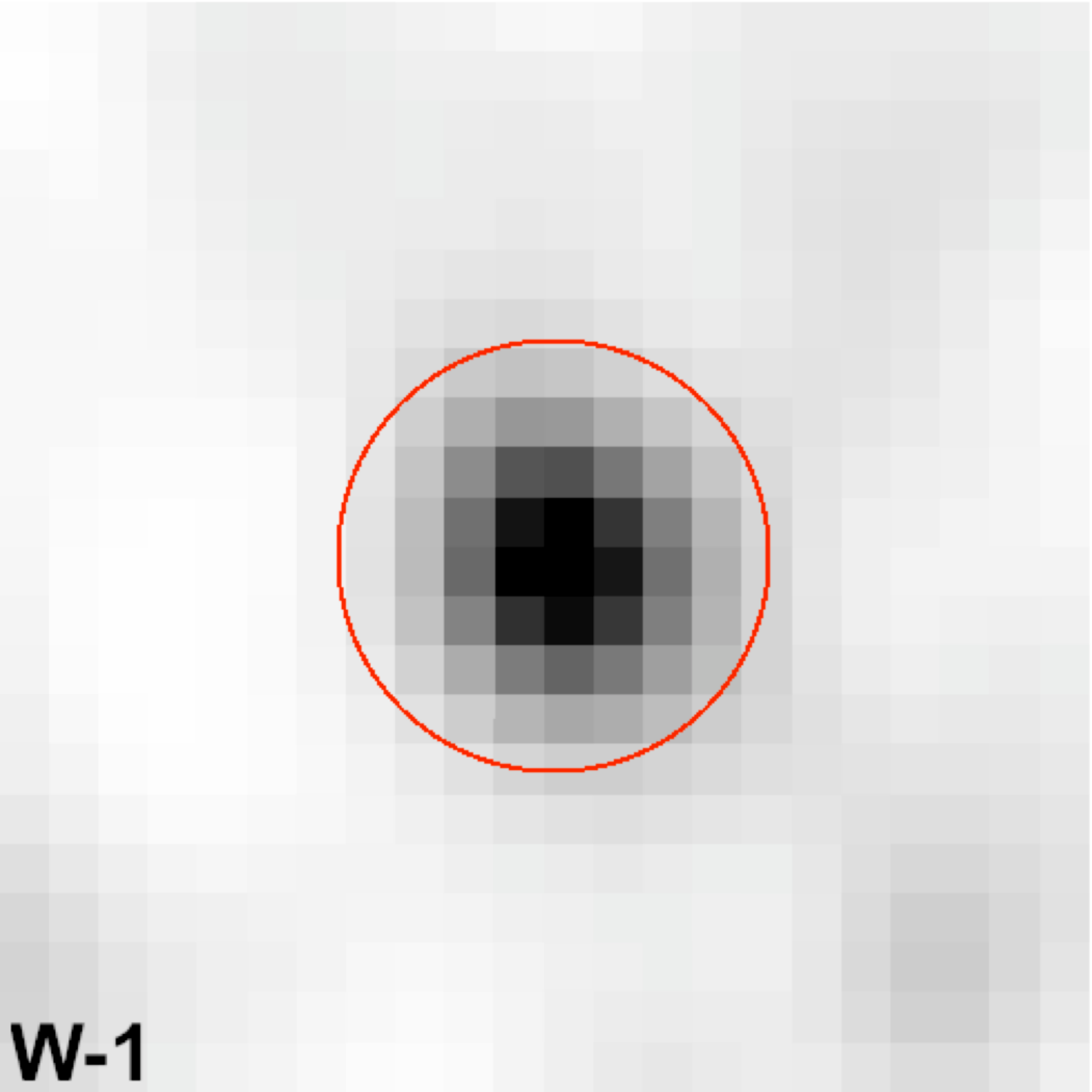}
\end{minipage}\hfill
\begin{minipage}{.3\textwidth}
\centering
\includegraphics[width=\linewidth]{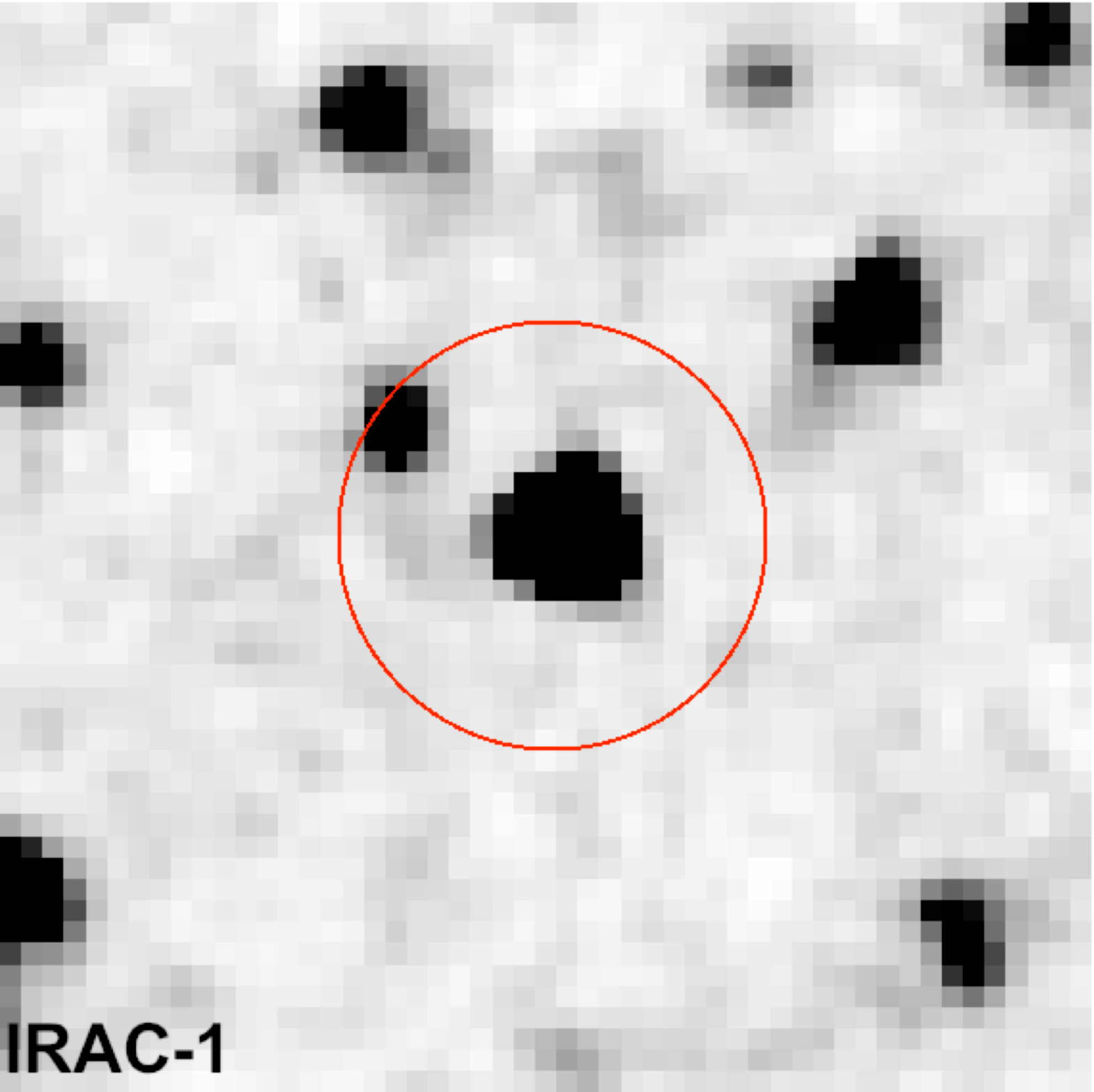}
\end{minipage}\hfill
\begin{minipage}{.3\textwidth}
\centering
\includegraphics[width=\linewidth]{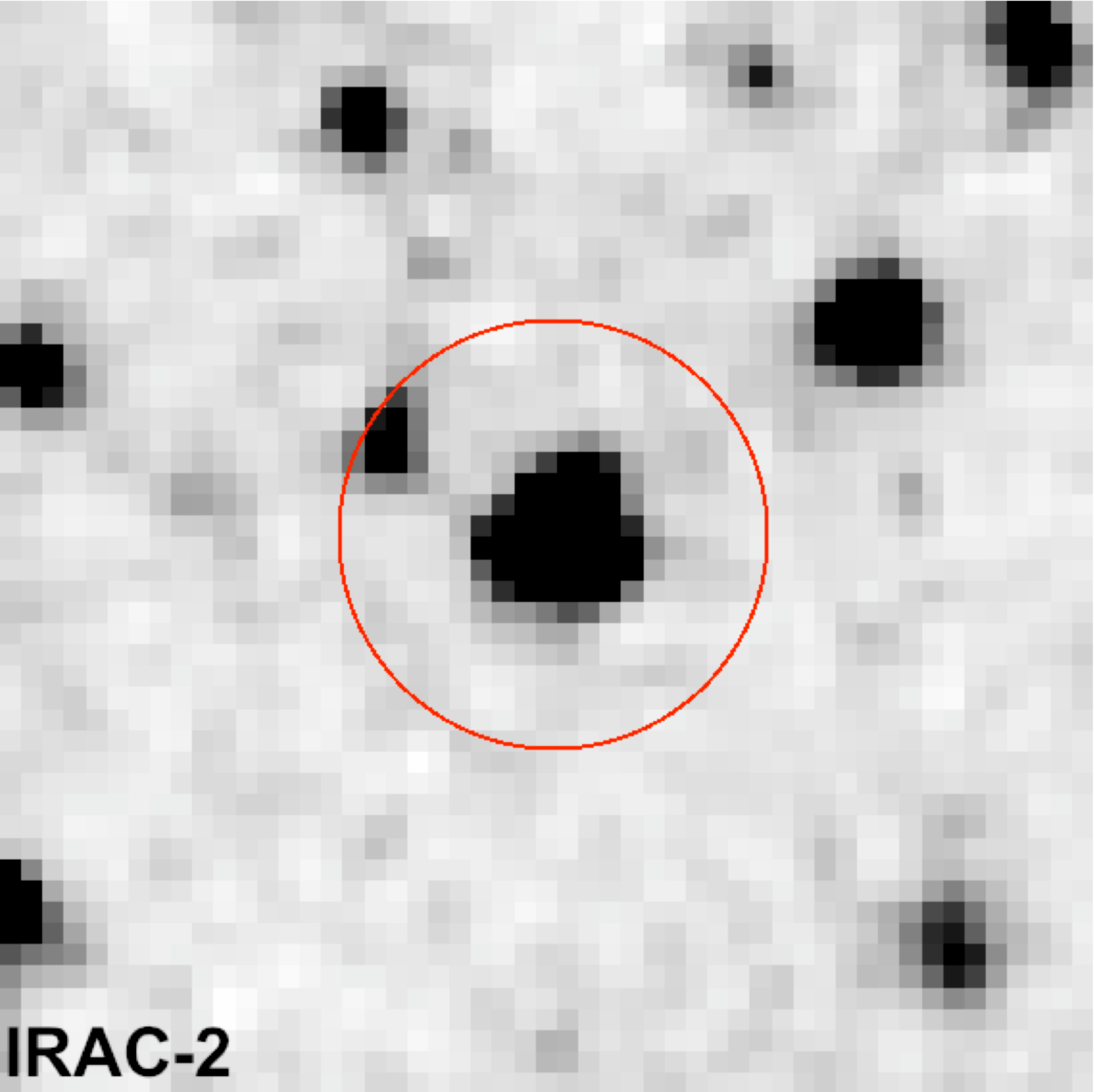}
\end{minipage}
\caption{Images for WD J0959$-$0200 in J, H and K bands (UKIRT, March 2014), W-1 band from ALLWISE as well as Spitzer/IRAC bands (February 2014). North is up and East is left. The field of view is 30{\farcs}0 by 30{\farcs}0. Red circles are centered on WD J0959$-$0200 with a radius of 6{\farcs}0, the resolution of WISE. In the IRAC images, there is a background source 5{\farcs}7 away from the target, which can affect the WISE results. The IRAC images are much deeper than the WISE image due to its higher sensitivity and longer exposure times.
}
\label{Fig: Image}
\end{figure}

Following data reduction procedures outlined in \citet{Farihi2010b} and \citet{XuJura2012}, we used MOPEX to create the final mosaic of 0{\farcs}6 pixel$^{-1}$, as shown in Figure \ref{Fig: Image}. In the IRAC bands, WD J0959$-$0200 is well separated from all nearby sources. However, there is a background source 5{\farcs}7 away, which could affect the WISE results (see section \ref{WISE}). We performed aperture photometry with PHOT in IRAF and the results are presented in Table \ref{Tab: Photometry}.

\begin{figure}[hp]
\centering
\begin{minipage}{.5\textwidth}
\centering
\includegraphics[width=\linewidth]{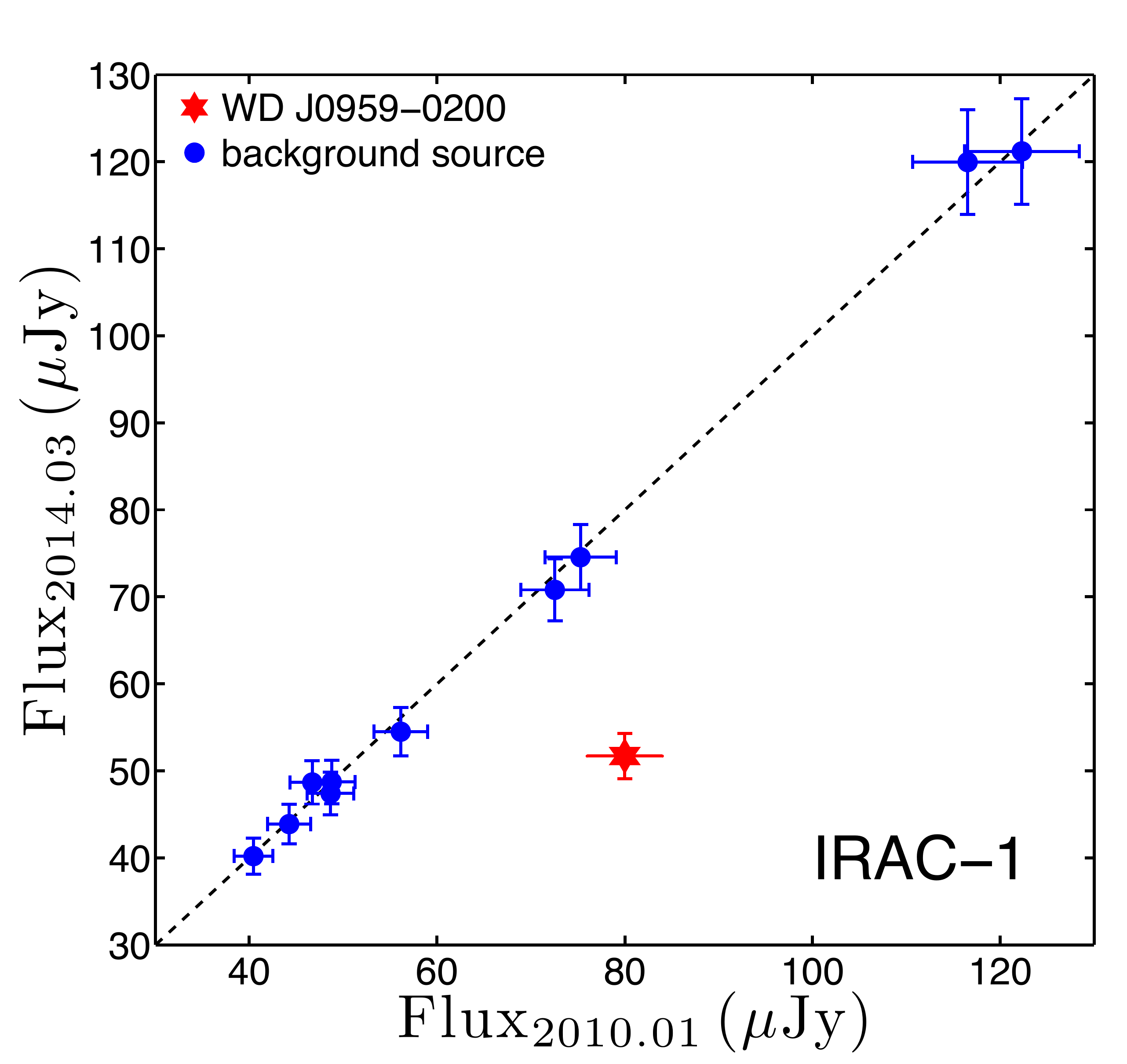}
\end{minipage}\hfill
\begin{minipage}{.5\textwidth}
\centering
\includegraphics[width=\linewidth]{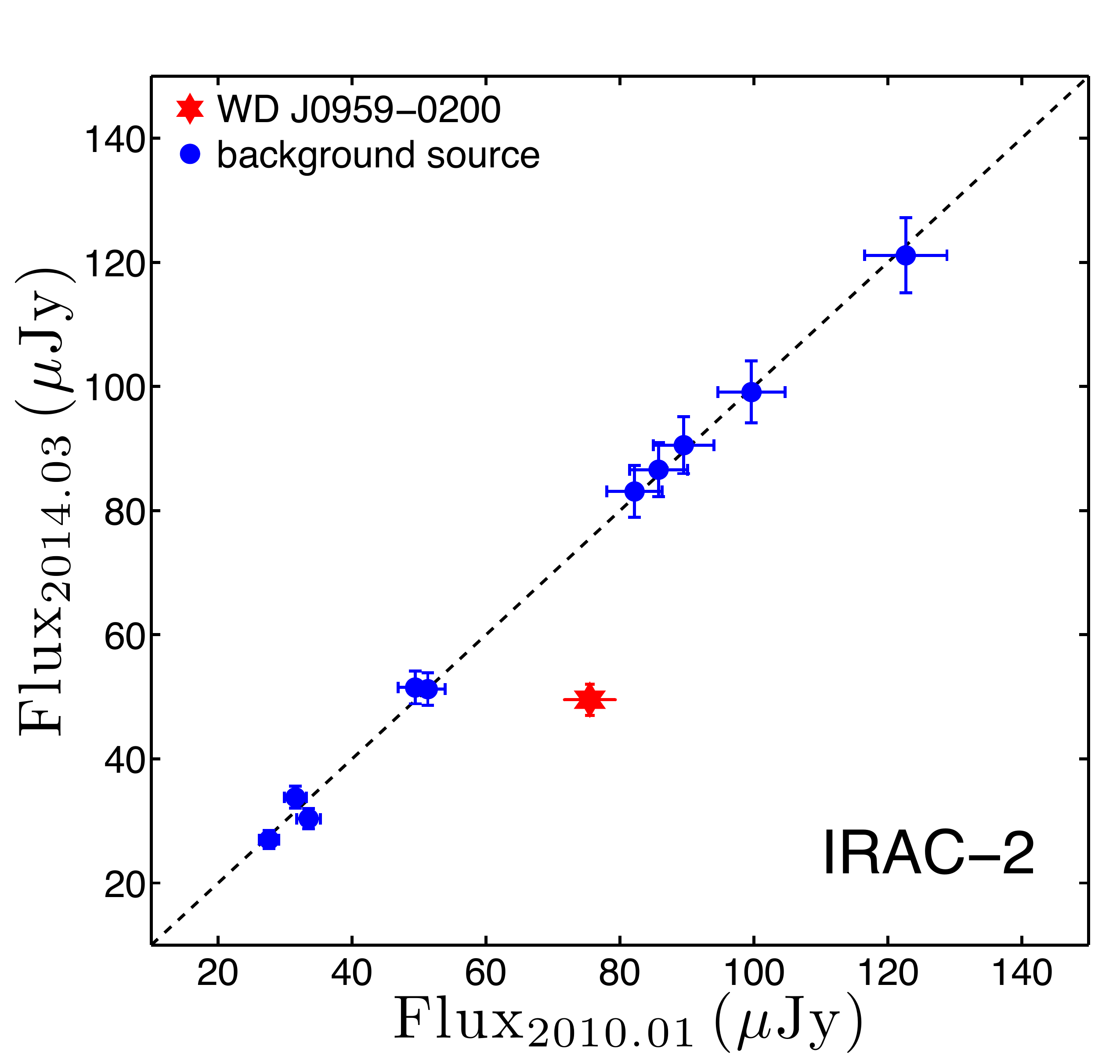}
\end{minipage}\hfill
\begin{minipage}{.5\textwidth}
\centering
\includegraphics[width=\linewidth]{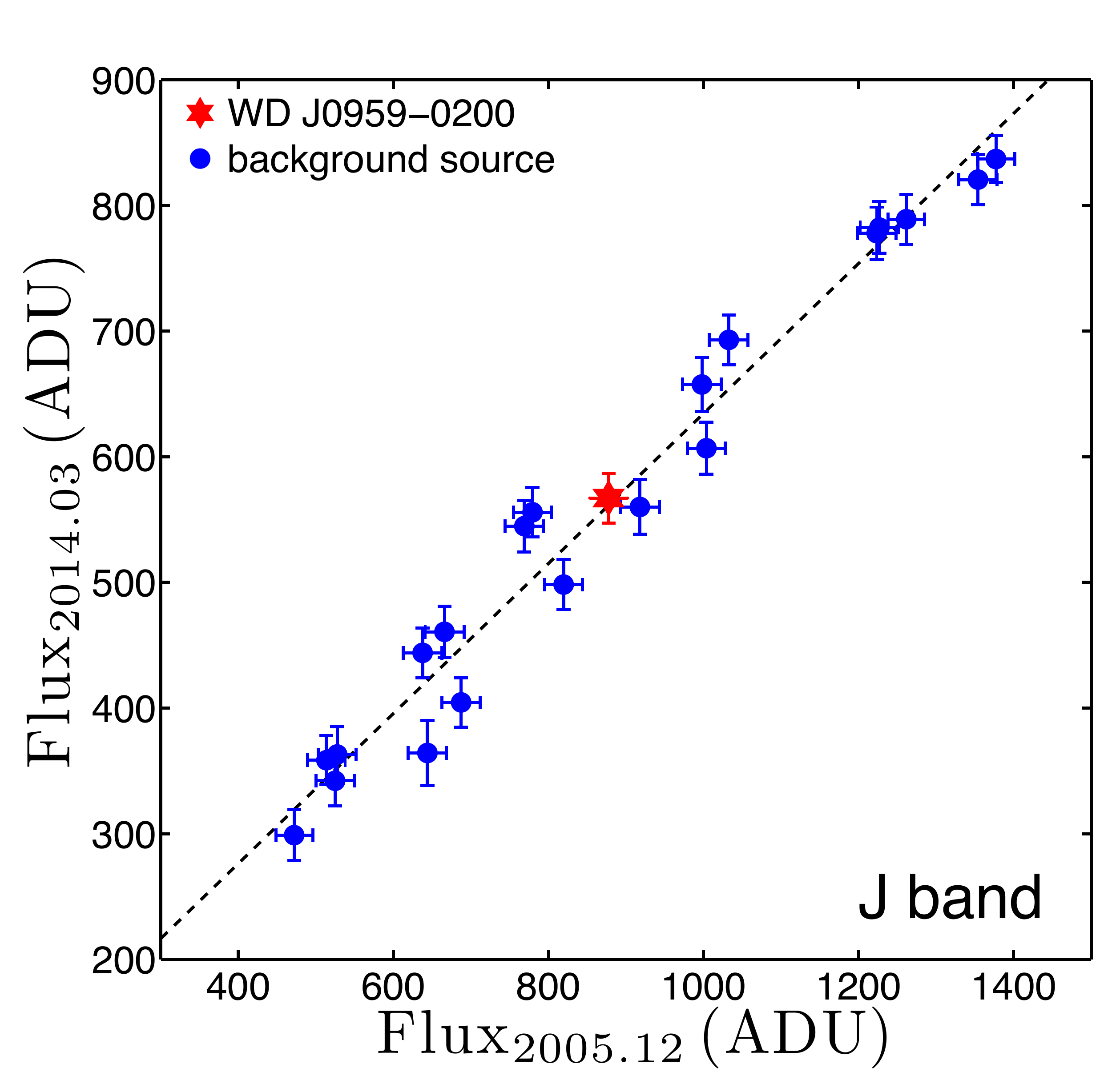}
\end{minipage}\hfill
\begin{minipage}{.5\textwidth}
\centering
\includegraphics[width=\linewidth]{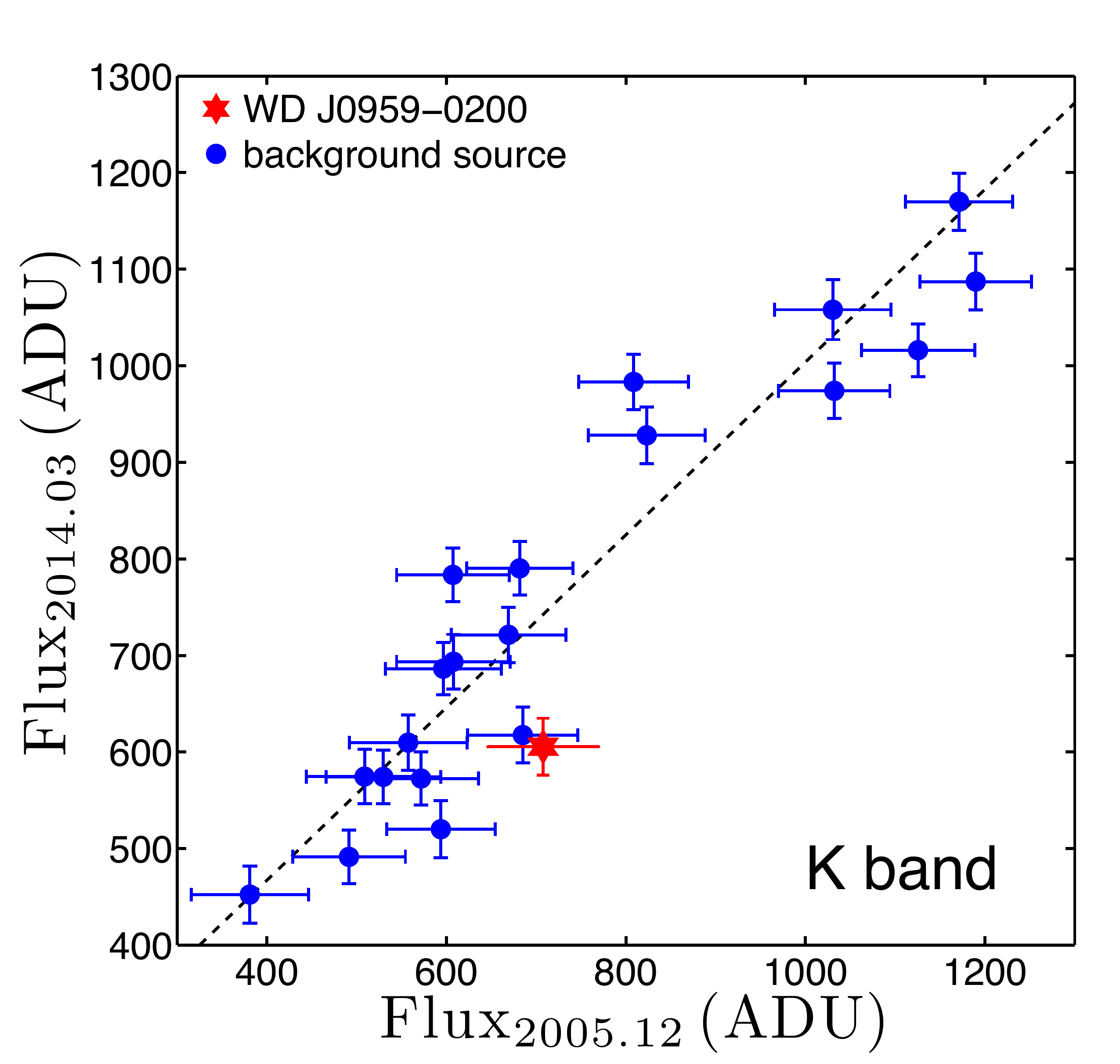}
\end{minipage}\hfill
\caption{Photometric measurements for WD J0959$-$0200 and nearby sources at different observing times, indicated by the subscripts. Upper panels: Spitzer/IRAC data. The dashed lines represent 1:1 correlations. The fluxes for WD J0959$-$0200 have decreased significantly in both IRAC bands. Lower panels: J and K band data from the UKIRT. The black dashed lines represent the least square fit to the 20 nearby sources. The J band flux has stayed the same while likely the K band flux has decreased, consistent with the IRAC results.
}
\label{Fig: comp}
\end{figure}

Relative to the 2010 data, we find that the flux for WD J0959$-$0200 decreased 35 $\pm$ 4 \% and 34 $\pm$ 4 \% in IRAC-1 and IRAC-2, respectively. For a sanity check, we also measured the fluxes of 10 nearby stars that have comparable fluxes and are in a clean field. As shown in Figure \ref{Fig: comp}, the standard deviation between the two epochs for the comparison targets is 1.3\% in IRAC-1 and 2.0\% in IRAC-2, much smaller than the observed flux differences for WD J0959$-$0200. The decrease of flux for WD J0959$-$0200 is real.

\subsection{UKIRT}

WD J0959$-$0200 was observed in 2005 with the Wide Field Camera (WFCAM) as part of the Large Area Survey (LAS) in the UKIDSS \citep{Lawrence2007}. Four broadband photometric observations at Y, J, H and K with an exposure time of 40 sec each were obtained. 

In 2014, we were awarded UKIRT DDT time to reobserve WD J0959$-$0200. A number of exposures were taken with 10 sec each. The total on target time was 150 sec, 450 sec and 1250 sec for J, H and K band respectively. No flux standard was taken. The raw data were processed by the pipeline from CASU (Cambridge Astronomical Survey Unit) and individual frames were combined in IRAF to produce the final mosaic, as shown in Figure \ref{Fig: Image}. We performed aperture photometry on the target as well as a few nearby sources. Because UKIDSS data covered the same surrounding stars, we used their published magnitudes to derive the flux calibration of the 2014 UKIRT data. The photometry results for WD J0959-0200 are listed in Table \ref{Tab: Photometry} and we see its K band flux might have decreased.

To further investigate the flux level for WD J0959$-$0200 in the two epochs, we performed aperture photometry on both the UKIDSS data and the 2014 UKIRT data. The flux unit was kept in analog to digital unit (ADU), which is directly from the raw data. This avoids introducing additional uncertainties from flux calibration because these stars are close to the limiting magnitude of the UKIDSS survey. A least square fit on the 20 nearby sources is used to find the conversion between the two epochs, as shown in Figure \ref{Fig: comp}. Relative to the flux in 2005, the J, H and K band fluxes for WD J0959$-$0200 in 2014 are 101.0 $\pm$ 4.5 \%, 99.1 $\pm$ 10.8 and 81.5 $\pm$ 8.1\%, respectively. Likely, the K band flux for WD J0959$-$0200 has decreased, consistent with the Spitzer results. 

\subsection{WISE \label{WISE}} 

The ALLWISE fluxes is 55 $\pm$ 6 $\mu$Jy for W1 and 49 $\pm$10 $\mu$Jy for W2. In addition, the IRAC images show that there is a background source within the WISE resolution of 6{\farcs}0 (see Figure \ref{Fig: Image}) and the measured fluxes are 9$\pm$2 $\mu$Jy and 8$\pm$2 $\mu$Jy for IRAC-1 and IRAC-2, respectively. We subtract those numbers from the ALLWISE data and list the corrected WISE fluxes in Table \ref{Tab: Photometry}.

\begin{table}[hptb]
\begin{center}
\caption{Multiwavelength photometry for WD J0959$-$0200}
\begin{tabular}{lllllccc}
\\
\hline \hline
UT Date & Instrument	& \multicolumn{3}{c}{Flux ($\mu$Jy)} & Ref\\
  \hline
2005 Dec 28 	& WFCAM	& J: 72$\pm$4	& H: 55$\pm$5	& K: 57$\pm$6	& (1) \\
2014 Mar 27	& WFCAM	& J: 73$\pm$3	& H: 54$\pm$4	& K: 46$\pm$3	& (2) \\
\\
2010 Jan 06	& IRAC$^b$	& IRAC-1:	 80$\pm$4& IRAC-2: 76$\pm$4	& & (3) \\
2010 May 18 \& Nov 24$^a$	& WISE$^b$	& W-1: 46$\pm$6$^c$	& W-2: 41$\pm$10$^c$	&	& (4) \\
2014 Feb 04	& IRAC$^b$	& IRAC-1: 52$\pm$3 & IRAC-2: 50$\pm$3 & & (2) \\

\hline
\label{Tab: Photometry}
\end{tabular}
\end{center}
{\bf Ref.} (1) UKIDSS; (2) measurement from this paper; (3) \citet{Farihi2012a}, our measurement also agrees with their values; (4) ALLWISE.\\
{\bf Notes.} \\
$^a$ WD J0959$-$0200 was observed by WISE at two different epochs but the intrinsic faintness of this star does not allow for reliable photometric measurement at each epoch. \\
$^b$ Central wavelengths for IRAC bands are 3.6 $\mu$m and 4.5 $\mu$m; for WISE bands, they are 3.4 $\mu$m and 4.5 $\mu$m.\\
$^c$ This is the corrected flux by subtracting the flux of the background source measured from IRAC, see section \ref{WISE} for details.
\end{table}

\section{Discussion}

The most stringent constraint on the duration of the infrared variability is between the first IRAC and WISE observations, which are 4-10 months apart. Luckily, optical spectroscopic observations with VLT/X-Shooter of WD J0959$-$0200 were obtained in June 2010 and March 2011, around the same time that WISE data were taken. No variations in the photospheric absorption lines or calcium triplet emission lines were reported. Given that WD J0959$-$0200 is directly accreting from the disk and the settling times of heavy elements are only days, there must have been little change in the physical conditions between the two epochs. Likely, the drop in the infrared luminosity has taken place in less than 5 months, between the first IRAC observation in January 2010 and the first X-Shooter observation in June 2010. However, it is more conservative to estimate a duration of the drop to be less than 10 months given the uncertainties in the individual measurements. The actual duration can be much shorter. 

\subsection{SED Fits}

The white dwarf spectrum is calculated using TLUSTY \& SYNSPEC \citep{HubenyLanz1995} and input parameters of T$_{wd}$ = 13,280 K and log g= 8.06 \citep{Farihi2012a}. To fit the SED of the dust disk, we use the opaque disk model described in \citet{Jura2003} with three free parameters, the inner disk radius R$_{in}$, outer disk radius R$_{out}$ and disk inclination i. The inner and outer disk radius is most sensitive to $\sim$ 3$\mu$m and 8$\mu$m flux, respectively while the disk inclination affects all the infrared flux. For the majority of dusty white dwarfs, the inner disk radius corresponds to a disk temperature of 1200 K, the characteristic sublimation temperature for silicates. The outer disk radius varies but is always within the tidal radius of the white dwarf.

\begin{figure}[hp]
\plotone{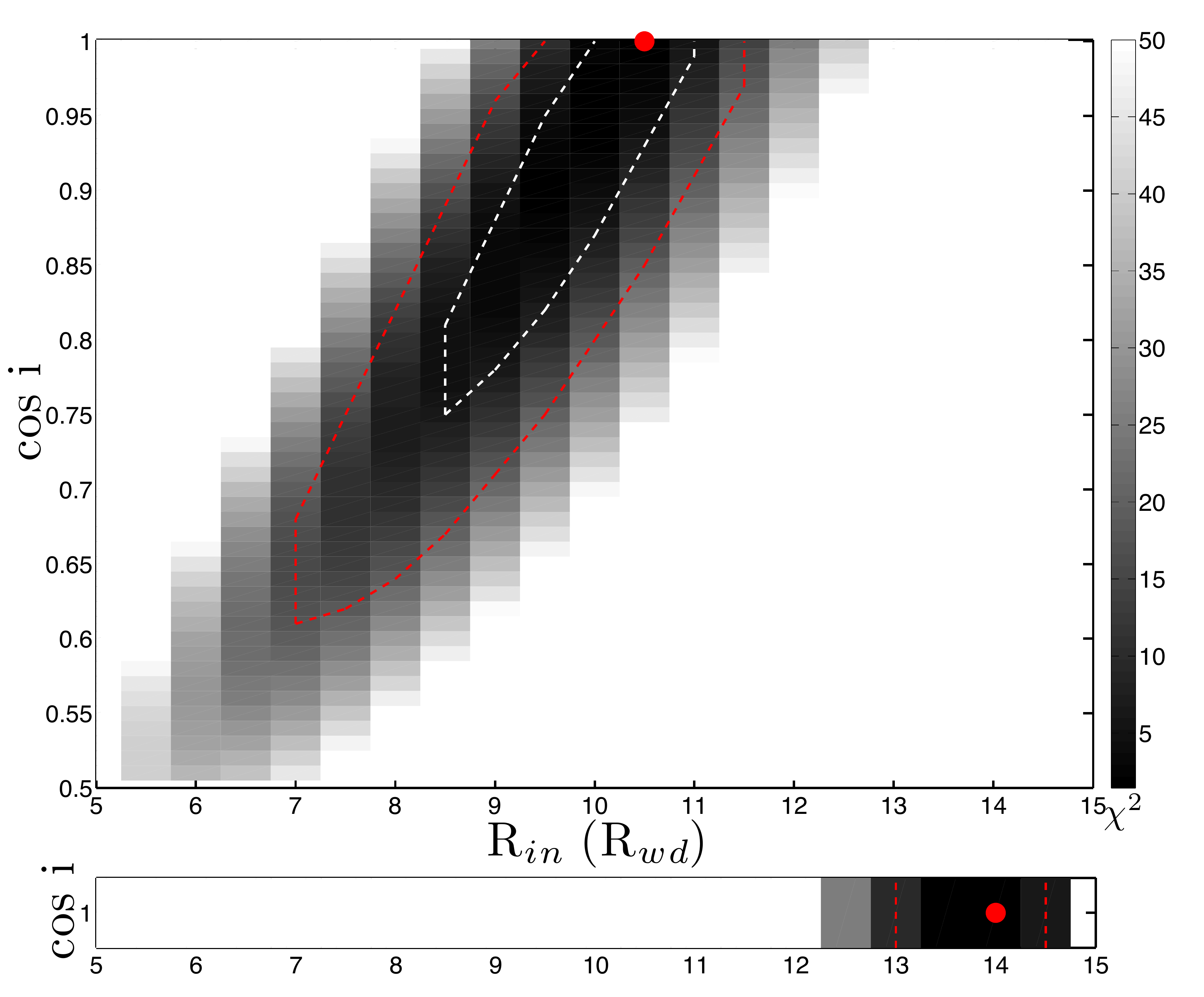}
\caption{The least $\chi^2$ fitting for the SED of WD J0959-0200. Top panel: a slice of the $\chi^2$ volume for pre March 2010 data with R$_{out}$ = 26.5 R$_{wd}$. The red dot is the least $\chi^2$ value of 1.42. The white dashed line is the 1$\sigma$ contour, which represents a disk inclination between 0 degrees and 41 degrees and an inner radius, R$_{in}$, between 8.5 R$_{wd}$ and 11 R$_{wd}$; the red dashed line is the 3$\sigma$ contour that has a disk inclination between 0 degrees and 52 degrees and R$_{in}$ between 7 R$_{wd}$ and 11.5 R$_{wd}$. A nearly face-on disk with a small inner radius is required to fit the SED. Bottom panel: $\chi^2$ fitting for the post March 2010 data with R$_{out}$ = 26.5 R$_{wd}$ and cos i=1. The red dot is the least $\chi^2$ value of 11.5. The red dashed line represents the 3$\sigma$ boundary and has  R$_{in}$ between 13.0R$_{wd}$ and 14.5R$_{wd}$.
}
\label{Fig: Contour}
\end{figure}

\begin{figure}[hp]
\plotone{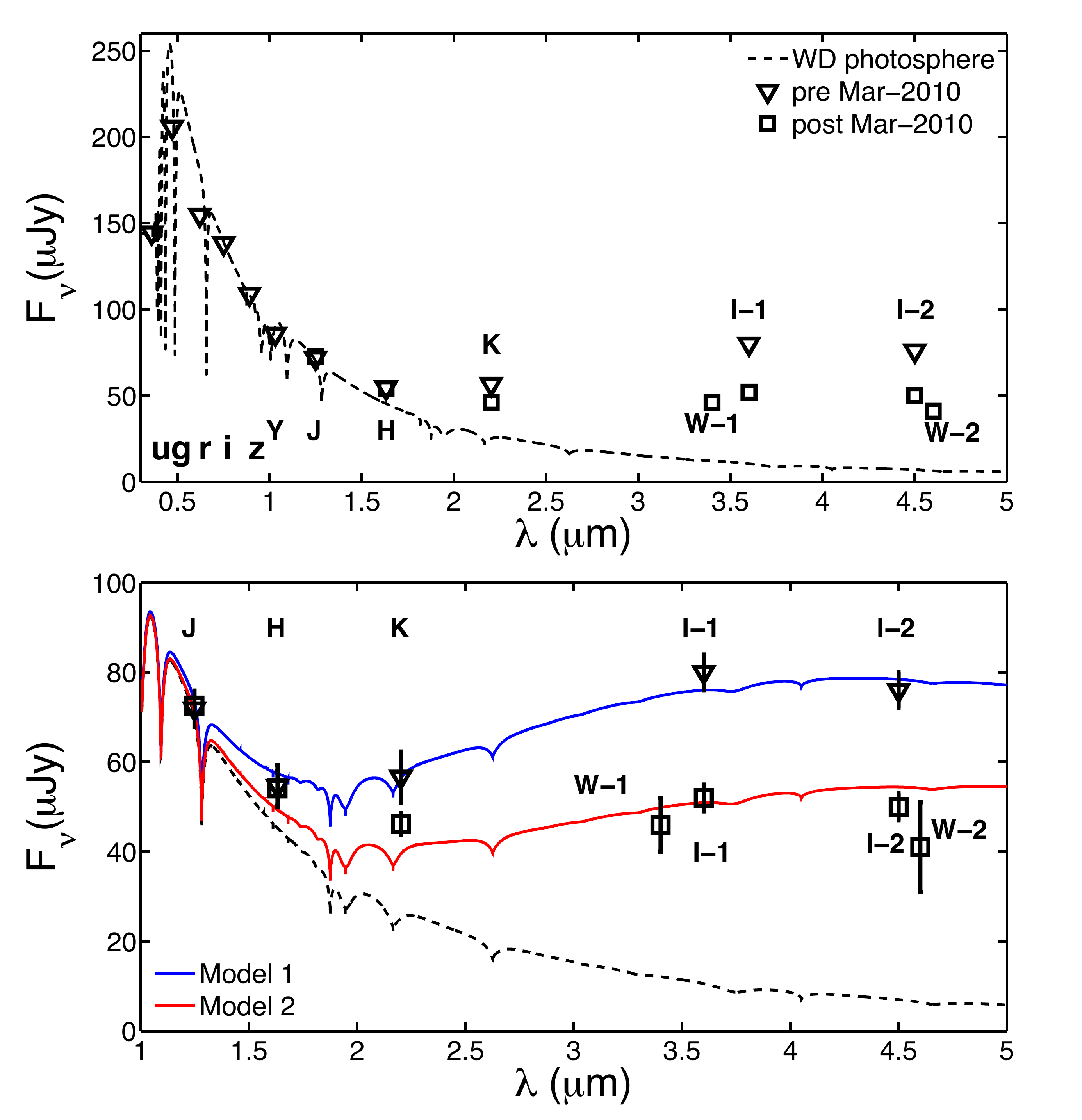}
\caption{SED fits for WD J0959$-$0200 with photometric data from the SDSS (ugriz), UKIRT (YJHK), WISE (W-1, W-2 = bands 1 and 2, respectively) and Spitzer/IRAC (I-1, I-2 = bands 1 and 2, respectively). The upper panel shows the entire SED and the lower panel is only for the infrared part. The black dashed line is the simulated spectrum for WD J0959$-$0200. The blue and red solid lines represent models that fit the pre Mar-2010 and post Mar-2010 data, respectively. The only difference between these two models is the inner disk radius, which has increased from 10.5R$_{wd}$ in Model 1 to 14R$_{wd}$ in Model 2.
} \label{Fig: SED}
\end{figure}

For the pre March 2010 data, we find the fractional luminosity L$_{IR}$/L$_{wd}$ = 0.03, the highest of all dusty white dwarfs \citep{Farihi2010b}. A  face-on disk with a small inner disk radius, correspondingly a high inner disk temperature, is needed to fit the SED. The best-fit disk parameters are R$_{in}$=10.5R$_{wd}$ (T$_{in}$ = 1545 K), R$_{out}$=26.5R$_{wd}$ (T$_{out}$ = 771 K), i = 0$^\circ$ and $\chi^2$ = 1.42 with 5 degrees of freedom from the fluxes of JHK bands and IRAC bands. The outer disk radius is not as well constrained due to the lack of data from longer wavelengths. A slice of the $\chi^2$ volume is shown in Figure \ref{Fig: Contour}. A less face-on disk with a even smaller inner radius can also fit the data; however, it would require a even higher inner disk temperature, which is harder to achieve. The SED with the best fit model is shown in Figure \ref{Fig: SED}. A caveat here is that there are 4 years between the UKIDSS and IRAC data and there could have been additional changes in the disk. However, the high fractional luminosity of the disk makes the quality of the fit insensitive to the likely range of the K band magnitude.

For the post March 2010 data, the fractional luminosity has dropped to 0.02, which is still on the high side for dusty white dwarfs. There is no unique fit to the near infrared data due to the degeneracy between the inner disk radius and disk inclination. Fortunately we can use the calcium triplet lines, which are much weaker and narrower than the other known cases; \citet{Farihi2012a} derived a nearly face-on disk with an inclination of 11$^\circ$. We set R$_{out}$=26.5R$_{wd}$, i = 0$^\circ$ -- the same as the model for pre March 2010 data -- and find the best fit model requires R$_{in}$=14R$_{wd}$ (T$_{in}$ = 1245 K). The $\chi^2$ fitting is shown in Figure \ref{Fig: Contour} and the SED is shown in Figure \ref{Fig: SED}. Even though the computed SED underestimates the K band flux, the model has a $\chi^2$ value of 11.5, which is within 3$\sigma$. With all the available data, an increase of the inner disk radius from 10.5$^{+1.0}_{-3.5}$ R$_{wd}$ to 14.0$^{+0.5}_{-1.0}$ R$_{wd}$ is the most plausible explanation for the drop of the infrared luminosity of WD J0959$-$0200.

\subsection{Interpretations}

What can cause such a change in the inner disk radius? We explore two scenarios and propose some experiments to test these hypotheses.

{\it A recent planetesimal impact}. \citet{Wyatt2014} suggest that tidal disruption of small planetesimals around white dwarfs can be nearly continuous. It is unlikely that the newcoming asteroid has the exact same orbital inclination as the pre-existing dust disk. \citet{Jura2008} showed that during an impact, the dust disk will be eroded by sputtering and the infrared flux will decrease within 600 yr, possibly much shorter. To increase the inner radius from 10.5R$_{wd}$ to 14R$_{wd}$, the disk has decreased 5 $\times$ 10$^{10}$ km$^2$ in surface area, much larger than the typical size of an asteroid in the solar system. For comparison, there exist dusty white dwarfs with a much larger inner hole, e.g. PG 1225-079 and G166-58, possibly also due to planetesimal impacts \citep{Farihi2010b}. Recently, tidal disruption of an entire planet by a white dwarf has been invoked to explain the light curve of a source in the center of a globular cluster \citep{DelSanto2014}.

{\it Instability in the circumstellar disk}. WD J0959$-$0200 belongs to the rare category of dusty white dwarfs which also display calcium triplet emission lines from an orbiting gaseous disk \citep{Farihi2012a}. To fit the pre Mar-2010 SED, even with a face-on disk, a high inner disk temperature of 1545 K is required, much higher than the characteristic temperature for rapid silicate sublimation\footnote{The majority of gas disk host white dwarfs require a similar high inner disk temperature \citep{Brinkworth2012}. The cause is unknown.}. \citet{RafikovGarmilla2012} and \citet{Metzger2012} found that this superheated inner disk region is unstable and interactions between the dust and gaseous material can cause rapid changes in the physical conditions. This can lead to a change in the viscosity of the gas and accretion rate onto the white dwarf. A flare is expected to occur \citep{BearSoker2013}, similar to what happens during dwarf novae outbursts [e.g. \citet{Lasota2001}]. To estimate the total mass loss, we approximate the average mass of heavy elements in a dusty helium white dwarfs' atmosphere 10$^{22}$ g as the disk mass. Increasing the inner disk radius from 10.5 R$_{wd}$ to 14 R$_{wd}$ corresponds to a total mass loss of  3.3 $\times$ 10$^{20}$ g. The derived mass accretion rate is at least $\sim$ 1.3 $\times$ 10$^{13}$ g s$^{-1}$ and an accretion luminosity of 1.5 $\times$ 10$^{30}$ erg s$^{-1}$, which is nearly 10\% of the bolometric luminosity of WD J0959$-$0200. The accretion luminosity is mostly like to be in the X-ray \citep{Jura2009b}, detectable with current X-ray facilities. In addition, there can also be optical transient events associated with the accretion, as seen in dwarf novae \citep{BearSoker2013}.

We estimate that $\sim$ 3\% of the mass of the disk was accreted onto the white dwarf in less than 300 days while the disk lifetime is estimated to be 10$^5$ - 10$^6$ yr. One possible implication is that the accretion occurs in short, intense bursts rather than in a steady state, as argued by \citet{Farihi2012b}. Future high precision, high cadence infrared photometric observations will be essential to disentangle different scenarios. If we can measure the photospheric abundance when the infrared flux is changing, we may have the opportunity of witnessing the specific composition of an incoming planetesimal. That is, the settling times of heavy elements can be quite short, e.g. a few days in WD J0959$-$0200, and we can imagine the observed abundance to change on a similarly short timescale.

\section{Conclusions}

We report the drop of infrared luminosity of a dusty white dwarf WD J0959$-$0200 in less than 300 days in 2010; the actual duration might be even shorter. Such a change can be caused by an increase in the inner disk radius. We provide two hypotheses to explain the change, including a recent planetesimal impact and instability in the circumstellar disk. Future observations in the time domain would greatly improve our knowledge of the fate of planetary systems around white dwarfs.

The authors thank an anonymous referee for helpful comments. S. Xu thanks R. Mostardi for helpful discussion on data reduction with IRAF. This work has been supported by the NSF. It is based in part on observations made with: (i) the Spitzer Space Telescope, which is operated by JPL, Caltech under a contract with NASA; (ii) WISE, which is a joint project of UCLA and JPL/Caltech, funded by NASA; (iii) data obtained as part of the UKIDSS. We also thank Richard Green for awarding us DDT time on the UKIRT and Tom Kerr for carrying out the observation. 

\end{CJK}
\bibliographystyle{apj}

\end{document}